\documentstyle[12pt,epsfig]{article}
\newfont{\Bbb}{msbm10 scaled 1200}     
\newcommand{\mathbb}[1]{\mbox{\Bbb #1}}

\def\lbldef#1#2{\expandafter\gdef\csname #1\endcsname {#2}}
\def\eqn#1#2{\lbldef{#1}{(\ref{#1})}%
\begin{equation} #2 \label{#1} \end{equation}}

\def\href#1#2{#2}

\newcommand{\beq}{\begin{equation}}
\newcommand{\eeq}{\end{equation}}

\newcommand{\ber}{\begin{eqnarray}}
\newcommand{\eer}{\end{eqnarray}}

\newcommand{\beqar}{\begin{eqnarray}}

\newcommand{\eeqar}{\end{eqnarray}}

\newcommand{\ba}{\begin{eqnarray}}
\newcommand{\ea}{\end{eqnarray}}

\newcommand{\dsl}
  {\kern.06em\hbox{\raise.15ex\hbox{$/$}\kern-.56em\hbox{$\partial$}}}

\newcommand{\eeqarr}{\end{eqnarray}}
\newcommand{\ZZ}{{\rm \kern 0.275em Z \kern -0.92em Z}\;}

\def\be{\begin{equation}}
\def\ee{\end{equation}}
\def\bea{\begin{eqnarray}}
\def\eea{\end{eqnarray}}

\begin{document}
\baselineskip=15.5pt
\pagestyle{plain}
\setcounter{page}{1}
\begin{titlepage}

\leftline{\tt hep-th/0604051}

\vskip -.8cm

\rightline{\small{\tt UB-ECM-PF 06/01}}
\rightline{\small{\tt KUL-TF-06/12}}

\begin{center}

\vskip 1.7 cm

{\LARGE Heavy hadron spectra from spin chains and strings}
\vskip .3cm

\vskip 1.5cm
\vspace{20pt}
{\large 
A.L. Cotrone $^{a}$, L. Martucci $^{b}$,}
{\large
J.M. Pons $^{a}$ and P. Talavera  $^{c}$
}
\vskip 1.2cm

\textit{$^a$ Departament ECM, Facultat de F\'isica, Universitat de Barcelona and \\ Institut
de Fisica d'Altes Energies, Diagonal 647, E-08028 Barcelona, Spain.}\\
\textit{$^b$ Institute for theoretical physics, K.U. Leuven,\\
Celestijnenlaan 200D, B-3001 Leuven, Belgium.}\\
\textit{$^c$ Departament de F\'isica y Enginyeria Nuclear,  Universitat Politecnica de Catalunya, Jordi Girona 1-3, E-08034 Barcelona, Spain.}

\vskip 0.5cm

\vspace{1cm}

{\bf Abstract}

\end{center}
We study the spectrum of hadronic states made up of very massive complex
scalar fields in a confining gauge theory admitting a supergravity dual background. 
We show that for a sub-sector of operators dual to 
certain spinning strings, the mass spectrum exhibits an integrable structure 
equal to the Heisenberg spin chain,
up to an overall factor. 
This result is compared with the corresponding string prediction.

\noindent

\end{titlepage}

\newpage


\section{Introduction}
\label{intro}

Since its appearance, the gauge/string theory correspondence has provided an alternative point of view in the study of many problems in quantum field theory. In particular, the existence of string theory backgrounds dual to {\em confining} gauge theories has allowed a deep insight in many issues related to confinement, chiral symmetry breaking and other important phenomena.

In this paper we will focus our attention on the study of hadronic states that are ubiquitous in the known confining theories that have a string theory dual.  The string theory analysis exhibits a precise and  universal structure for their mass spectrum, that has been analyzed by considering some particular limit in which the string theory becomes soluble or, more generally, studying some semiclassical string configurations dual to the hadrons  \cite{strassler,bcm,sonne,noi4}.  A pure field theory derivation of this hadronic spectrum, matching the string theory semiclassical results, would be important since it would provide a non-trivial check of the correspondence for non-supersymmetric (i.e. non-protected) observables. The aim of this work is to pursue such a field-theoretical study and to show that some important features of the spectrum derived from string theory can be indeed reproduced.

One of the key problems in all the checks of the gauge/string theory correspondence is that the supergravity description covers the strong coupling regime of the dual field theories. Then any perturbative field theory reproduction of supergravity  results seems to be hopeless, unless one considers only sectors protected by  supersymmetry. However, for the (conformal) AdS/CFT case, in the last years  (starting from \cite{bmn}) there has been a great improvement of the understanding of the correspondence in non-BPS sectors of the theory. In particular, in ${\cal N}=4$ SYM, a lot of attention has been focused on  certain (non-BPS) composite operators of the form Tr$[\Phi_{A_1}^{J_1}...\Phi_{A_k}^{J_k}]$, where $\Phi_A$ are the scalars of the theory. In \cite{minzar} it was shown that the one-loop matrix for their anomalous dimension  is given by an $SO(6)$ integrable spin chain.
This result has made possible the calculation, based on spin chains and using the Bethe ansatz techniques, of the anomalous dimensions of various such operators in the limit of a large number $J\equiv J_1+...+J_k$ of constituents. The large $J$ limit allows to organize the anomalous dimension spectrum in an expansion in $\lambda/J^2$, where $\lambda$ is the 't Hooft coupling. The interest of this limit is that the dual objects in string theory admit a semiclassical description. These objects correspond to spinning and/or pulsating string solitons with large quantum numbers on the five-sphere of $AdS_5 \times S^5$ \cite{gubser}. A crucial point is that the $\alpha'$ sigma-model corrections to the classical string solutions seem to be subleading at large $J$. Since also these classical solutions admit an expansion in $\lambda/J^2$, even if the solutions are {\sl a priori} valid at large $\lambda$, in the large $J$ limit one can try and extrapolate the results at small $\lambda$ and compare them with the field theory results, finding agreement.

Our aim is to figure out whether the recent developments in the conformal case have an analog for confining gauge theories, where the correspondence is between the {\sl physical mass} of  certain hadronic bound states --and their excitations-- and the energy of dual semiclassical stringy configurations. 
The hadrons are created by operators of the form Tr$[\Phi_{A_1}^{J_1}...\Phi_{A_k}^{J_k}]$, where now the adjoint scalar fields correspond to massive particles.
Our approach mimics in a certain sense the philosophy of \cite{Beisert:2004fv} focusing our attention just to the first excited states in a sub-sector with some fixed quantum numbers (for a review of integrability in Yang-Mills theories, see \cite{Belitsky:2004cz}). This allows to find a parameter window where the spectrum can partially fit an integrable spin chain.

Since we are talking about confining gauge theories,  the situation is much more complicated than in the even difficult conformal case, and at a first sight it could seem impossible  to obtain the spectrum of these hadrons from a standard perturbative field theory calculation. 
Indeed, in the supergravity regime the dual field theory is strongly
coupled in the UV  (in other words, it is not asymptotically free),
the mass of the adjoint matter is of the same order of the
confining scale and any perturbative field theory calculation breaks down.
However, we can think to continously deform the bare coupling constant until it becomes small enough. 
In this regime,
opposite to the one valid in supergravity, the
confining theory is weakly coupled in the UV and  the mass of the
adjoint scalars becomes much larger than the confining scale.
We can then consider hadronic states made up of these scalar fields. 
Since their mass scale is large one can work at weak coupling and reduce the problem to a perturbative non-relativistic quantum mechanical one \cite{witten,brambilla}. 
  
 As we are going to show in  this paper, this non-relativistic limit allows to use an effective reduced Hamiltonian characterized by a flavor dependent term which is identical to a standard spin chain. The resulting spectrum then exhibits the typical integrable structure obtained also from the dual supergravity picture. The only possible difference is in an overall factor, still undetermined. 
This is not surprising
 since in general the states we are considering are not supersymmetric
 (even in any asymptotic sense). Thus, the perturbative (and probably
also non-perturbative)
  corrections as we deform the bare coupling are expected to be relevant
  in both the string theory and field theory calculations.
However, our result suggests how they can be nevertheless kept under control, obtaining a non-trivial match between the two dual descriptions.


\section{The field theory setup}

To be specific, we will restrict our attention to the
confining theory 
whose supergravity dual is known as the Witten model of YM
in four dimensions \cite{wym}. 
We nevertheless expect the 
results to go through in other confining models in every dimension.

The starting point is a compactification of the  
${\cal N}=2$ five-dimensional SYM theory, which in turn may be obtained by dimensional 
reduction of $10d$ SYM on a torus.     
From the dual gravity side, one starts from a stack of M5 branes in $11d$ compactified 
on a circle to give D4 branes in $10d$ and hosting a $5d$ SYM living on them. 
The radius of the 
circle
sets the scale of the $5d$ YM coupling
$g^2_5$. 
The D4 branes are then compactified on a thermal $S^1$ imposing 
anti-periodic boundary conditions for the fermions. 
This breaks automatically 
supersymmetry and provides masses to the scalars at one-loop level. 
Below the thermal 
compactification scale the theory is effectively a $4d$ YM, with massive 
 matter in the adjoint representation.

For the $4d$ theory, the compactification scale gives the order of the mass of all the fermions 
$m_F\simeq 1/R_5$, which sets essentially a lower bound for the KK masses. 
$m_F$ plays the role of an UV cut-off, that we take very large compared to the YM scale, 
$\Lambda_ {YM}\ll m_F$. The coupling at the UV scale is 
$g^2\simeq g^2_5m_F$ which sets the 't Hooft coupling at this point as 
$\lambda \equiv \lambda(m_F) \simeq g^2_5m_FN$. 
The limit $\lambda \ll 1$ is then reachable for $g^2_5  \ll 1/(m_FN)$.

The scalars we are interested in are massless at tree level but get masses at one loop. The scale of this mass
is $m^2\simeq \lambda m_F^2$. In fact there appears a hierarchy of scales 
$\Lambda_ {YM}\ll m \ll m_F$ because $\Lambda_ {YM} \simeq m_F e^{-1/\lambda}$ and 
we are taking $\lambda \ll 1$. 
Thus it is possible to work in the non-relativistic limit at energies $E\ll m$, keeping decoupled the KK modes and being still at $E\gg \Lambda_ {YM}$.
The latter condition, together with the fact that $\lambda \ll 1$ and that, if we are away from $\Lambda_ {YM}$, the running coupling varies very slowly, allows to keep the latter perturbative at the scale we work, $\lambda(E) \ll 1$.
This means that for $m_F$ sufficiently large (that is, $\lambda$ sufficiently small) we can keep $\lambda(E)\simeq \lambda$. 
It also means on the other hand that, 
even the theory being non supersymmetric, the different couplings are all essentially the same,
$\lambda$, the corrections being subleading.

The outcome of this discussion is that, at the energy scale given by the (one-loop renormalized) mass $m$ of the adjoint scalars, the resulting effective four-dimensional theory contains only bosonic degrees of freedom given by the  gauge fields $A_\mu$ and the ``light" scalars $\Phi_A$ (the zero-modes of the KK scalar tower, the ones that get mass only at loop level). 

In fact, if in the reduction from five to four dimensions one does not put anti-periodic boundary conditions for fermions, the low-energy lagrangian would be just the one for ${\cal N}=4$ SYM, being all the KK modes decoupled at low-energy \cite{wym}\footnote{One can find the relevant formulas for the reduction for example in \cite{hebecker}.}.
In our case, on top of this, one has the explicit mass term for the fermions, given by the anti-periodic boundary conditions, so that also these modes ``live'' at high-energy and are decoupled from the low-energy theory and so do not enter the low-energy lagrangian.
Moreover, being the supersymmetry broken, the zero-modes of the scalars are no more protected against quantum correction to their mass, as already mentioned.
As a result, the low-energy lagrangian contains the mass term for the scalars.
One may be concerned with the fact that, having no more supersymmetry, in the low-energy regime one should have more interaction terms for the scalars that those of ${\cal N}=4$ SYM.
But if one works, as we will do, at the leading order in $\lambda$ and in the non-relativistic limit, all these other terms are subleading. In fact, whatever term will be generated, it will be either higher order in $\lambda$ or suppressed in powers of $1/m_F$ w.r.t. the bosonic ${\cal N}=4$ SYM interactions.
For example, other quartic terms, being generated at loops, would be subleading in $\lambda$  w.r.t. the ${\cal N}=4$ SYM quartic interaction of the scalars.
Analogously, interactions involving scalars and fermions at high-energy, at low-energy produce interaction terms for the scalars that are higher orders in $\lambda$, with the only exception of the mass term, that is order $\lambda$.
This can be easily seen from the fact that the lowest order diagram involving fermions running in loops, for external scalar states, is just the one loop mass term correction to the scalar propagator: all other pairs of internal fermion lines, generated by an ingoing scalar by the Yukawa coupling, that are not reconnecting to an outgoing scalar, must be re-absorbed in more vertexes, thus including higher powers of the coupling.   
Finally, higher dimension interactions will come with powers of $1/m_F$ and so will not be relevant in the non-relativistic limit.
The only term that is generated at loop level and that is relevant is the mass term for the scalars.
We will comment more about these points in the next sections, after discussing  the hadrons we want to study.

In the following we will study the spectrum of bound states involving only four of the (six) scalars, transforming in a flavor $SO(4)$ symmetry group and coming from the dimensional reduction of the hypermultiplet living on the higher dimensional D4 or M5-branes. Then, the effective action at the scale $m$ can be restricted to be of the form
\eqn{denlag}{
S = \int d^4 x\,\, {\rm Tr}\left[ -{1\over 2} D_\mu \Phi_A  D^\mu \Phi_A +{1\over 2} m^2 \Phi_A  \Phi_A 
- {1\over  4 g^2}F_{\mu\nu}F^{\mu\nu} + {g^2\over 4} [\Phi_A,\Phi_B]  [\Phi^A,\Phi^B]
\right]\,,\, 
}
where as usual $D_\mu = \partial_\mu + i \left[A_\mu,\cdot\right]\,$ and $F_{\mu\nu} 
= \partial_\mu A_\nu -\partial_\nu A_\mu+ i \left[A_\mu,A_\nu\right]$. 
This action is enough for our purpose, since it contains the ``light'' fields and their leading interaction terms in $\lambda$ and $1/m_F$.

In the next section we will further discuss the validity of the above action for the calculation we are interested in, that considers only  hadrons generated by operators of the symbolic form ${\rm Tr} [Z^{J_1}W^{J_2}]$, where we have introduced the complex scalars $Z=\Phi_1+i\Phi_2$ and $W=\Phi_3+i\Phi_4$; 
in other words, we restrict ourselves to the $SU(2)$ sector. 

Let us now pause for a moment and clarify a bit the procedure we are going to follow for our calculation.
Our goal is to compute the flavor dependent part of the mass spectrum of hadrons created by the operators ${\rm Tr} [Z^{J_1}W^{J_2}]$, at leading order in the coupling $\lambda$ and in a particular regime of parameters. 
That is, we are going to work at energies $E\ll m$, where the theory is non-relativistic and the problem become manageable.
In fact, at leading order in the small coupling $\lambda$ all that is needed is just the potential given by the non-relativistic limit of the tree level interaction terms in the effective lagrangian (\ref{denlag}).
This is a completely standard procedure, employed for example in \cite{witten} in the discussion of the baryons in the large $N$ limit, or in phenomenological description of heavy quarkonium, see for example \cite{brambilla}.

The procedure is distinct from the ways the spin chain is derived in \cite{minzar} and following literature.
There, one is concerned with the one-loop matrix of anomalous dimensions (in conformal theories) of the operators ${\rm Tr} [Z^{J_1}W^{J_2}]$.
At one loop, one has contributions from gluon exchange, scalar interactions and self-energy diagrams, that combine to give 
the spin chain.
Instead, here the aim is to calculate a mass spectrum in a confining theory, reducing the problem to a quantum mechanical one, calculating the potential at tree level.
Once we have identified the effective Lagrangian (\ref{denlag}), there are no further one-loop effects to take into account. All we have to do  is write down the relevant  non-relativistic Hamiltonian, computing the tree level interaction terms for the scalars and taking the non-relativistic limit. 
Remarkably, even if the problem addressed here is conceptually and a priori
quite different from the one considered in \cite{minzar}, we will see how the
same spin-chain structure emerges also in the present context.

As we have discussed, in the field theory regime (as opposite to the supergravity one) at energies $E\ll m$ the dynamics can be approximated by a non-relativistic Schr\"odinger equation.  
The only two dynamical contributions at tree level that will be relevant  are the exchange of a gluon between 
two nearest scalars and a vertex of four scalars. 
The former yields the Coulomb potential. Taking the spatial momenta of the external 
incoming (outgoing) particles, $\vec p$ ($\vec p^\prime$), to be subleading with respect to their 
 mass, this contribution amounts to
\begin{equation}
 \frac{4i\lambda m^2}{N|\vec{p}-\vec{p}'|^2}\delta_i^j\delta_k^l\,,
\end{equation}
where $\delta$-factors concern flavor indices, telling whether the scalars are $Z$'s or $W$'s (i.e. $i=1$
 correspond to $Z$ and $i=2$ corresponds to $W$), and the explicit dependence on the color indices has been suppressed.
The explicit expression of the resulting  potential between two particles,
in the space of positions, 
reads 
of course
\be\label{coulomb}
V_{{\rm Coulomb}}(\vec x- \vec x')^{jj'}_{ii'}=- \frac{\lambda}{ |\vec{x}-\vec{x}'|}\delta_i^j\delta_{i'}^{j'}\,.
\ee
Let us remind the fact that, even if the theory is confining, the reduction of the potential to the Coulombic one in the non-relativistic limit is a very general phenomenon, in the large $N$ limit \cite{witten} as in QCD \cite{brambilla}.
It is due to the energy regime considered. 
As said, we work at an energy scale $E \gg \Lambda_{YM}$. 
This means that we probe short distances compared to the inverse dynamical scale, so the linear behavior of the confining potential is not relevant and one is left just with the short distance Coulomb factor.
In other words, at very high energy the non-perturbative effects driving the linear confinement are not the main contributions to the potential. 

The second contribution to the hadron dynamics is 
the quartic coupling of two adjacent scalars. These interactions are point-like, with the 
space-time contribution being just constant while the flavor one coincides with the equivalent 
of ${\cal N}=4$ SYM \cite{minzar} (again suppressing the color dependence),
\beq\label{quartic}
V_{{\rm quartic}}(\vec x- \vec x')^{jj'}_{ii'}=\frac{\lambda}{(2m)^2}\delta^3(\vec x-\vec x')(\delta_{i}^{j}\delta_{i'}^{j'}
-2\delta_{i}^{j'}\delta_{i'}^{j})\,.\eeq
In order to simplify notation, we have omitted  the precise color structure in (\ref{coulomb}) and (\ref{quartic}) since, as it is discussed  in the appendix, in the large $N$ limit it will not play any role in the effective Hamiltonian for the states we are interested in (see (\ref{ham}) below).

\section{Hadron masses from a spin chain}
\label{sec3}

We start by discussing more in detail the hadronic bound states we want to study and some general features of the effective theory that should describe them.

In the asymptotically free region each $\Phi_A$ is associated with a creation operator  
$\alpha_A(k)^\dagger=\alpha_A^a(k)^\dagger t_a\,,$ where $\left\{ t_a \right\} $ are a basis 
of hermitian matrices for $u(N)$.
A generic (non-normalized) {\emph{colorful}} state built up of $J$ components can be obtained by acting on 
the vacuum
\bea\label{colored}
\vert k_1,A_1,a_1;\ldots; k_J,A_J,a_J\rangle=\alpha_{A_1}^{a_1}(k_1)^\dagger\cdots 
\alpha_{A_J}^{a_J}(k_J)^\dagger|0\rangle\ .
\eea  
Among all the previous states, nature dictates that hadron must be colorless, 
and hence the physical spectrum must only contain the latter type. Tracing out the color operator 
we obtain an effective theory of {\em colorless} particles characterized by their momenta 
and flavors. The reduced Hilbert space ${\cal H}_{{\rm Hadrons}}$ is generated by states of the form 
given by 
\bea\label{colorless}
\vert \vert  k_1,A_1;\ldots; k_J,A_J\rangle\rangle={\cal N} {\rm Tr}(t_{a_1}\cdots 
t_{a_J})|k_1,A_1,a_1;\ldots; k_J,A_J,a_J\rangle\,,
\eea
where ${\cal N}$ is the appropriate normalization constant.    
Note that, even if the states  (\ref{colored}) (and then (\ref{colorless})) are by construction totally symmetric under the ``exchange'' of two elementary colored constituents, the hadronic states (\ref{colorless}) 
 are {\em not} totally symmetric under the exchange of the effective {\em colorless} particles as in the case of a spin chain.  
 This difference can be traced back to the identification of these kind of hadrons with operators of the form  ${\rm Tr}[\Phi_{A_1}\cdots\Phi_{A_J}](x)$. Obviously, any  reordering in the constituents, except for cyclicity, corresponds to a different hadron. 
Notice also that the spin chain under consideration is closed, an unavoidable requirement to map the 
solution of the mass spectrum to the diagonalization of a spin $1\over 2$ ferromagnetic Heisenberg chain.

As a consequence of this discussion, in the final effective quantum mechanical system for the  colorless 
hadrons, the effective colorless  constituent particles are distinguishable, just like in an ordinary chain of 
harmonic oscillators, and like the states (\ref{colorless}) the resulting effective Hamiltonian is required to respect only the cyclic symmetry.

Before proceeding it is worth to bear in mind that the theory under inspection is not conformal 
invariant and that  we must pay special attention to the stability of the would-be hadrons 
under consideration. In order to discuss this problem, let us stress that we will eventually focus our attention to hadronic states associated to operators of the form ${\rm Tr}[ZZWZWWWZ....](x)$. 
These operators will contain $J_1$ fields $Z$ and $J_2$ fields $W$ ($J\equiv J_1+J_2$), where $Z$ ($W$) is charged under the first (second) $U(1)$ factor inside the flavor $SO(4)$ symmetry group.

Let us recall that, being the theory a non-supersymmetric circle reduction of a $5d$ SYM, in $4d$ 
one expects a whole bunch of interaction terms of the scalars above.
But, to begin with, the fermion mass $m_F$ can be tuned to make it heavy enough to be integrated out in external legs. 
As a consequence one can not obtain any hadron decaying to another species plus some
fermionic states. 
As already mentioned, it is precisely the fermionic radiative corrections that are responsible for 
the existence of mass for our scalars. This fact can lead to a next-to-leading scalar mixing 
operators, signaling the non-fermionic decay of the hadrons into other hadronic species. 
One can verify that up to including one-loop graphs the fermion 
contribution to any bosonic scalar mixing operator is vanishing. Thus the mass matrix 
of the scalars is still diagonal in the original basis. 

Moreover, the mixing of 
the zero modes of the scalars under consideration with the other KK modes is suppressed, 
since the former get mass at loops --we are assuming the coupling very small--, while 
the latter have a large tree-level mass. Contributions to interactions
between the scalars other that gluon exchange and the quartic vertex either
are of higher order in $\lambda$ or are suppressed by mass factors $1/m_F$.   
We can argue that the dimensional reduction does not affect the $5d$ $SU(2)_R$ 
symmetry among the two complex scalars constituent of the hadrons.
One can thus see that the operators with only $Z$ and $W$ insertions mix only among 
themselves, for symmetry reasons, in the non-relativistic limit. For example, one cannot generate extra uncharged $Z/\bar Z$ pairs for energetic reasons, and by charge conservation the number of $Z$ and $W$ must be conserved.
This rules out the mixing with the other two real ``light'' scalars present in the theory, that are uncharged with respect to $SU(2)_R$.

Once the stability issue is under control, we must consider the question of statistics. According to our 
previous discussion the resulting hadrons will not present a symmetric wave function with 
respect to the exchange of constituent particles. 
Only cyclic permutations will enjoy this condition.  

Another delicate point is that without any further input the perturbative theory derived 
from (\ref{denlag})  becomes intractable even at tree level. To appreciate the problem, 
one may look at the lowest perturbative correction to the free propagation of a $J$ scalar 
state where one of the constituents is coupled to the other $(J-1)$. Note however that nothing of 
this sort occurs in the large $N$ limit, where there are, on the contrary, some drastic 
simplifications due to the fact that each constituent just interacts with its two nearest 
neighbors, being the rest of the interactions subleading.

From the discussion presented in the previous section and formulae (\ref{coulomb}),  (\ref{quartic}), we can now write the effective Hamiltonian describing the hadrons we are interested in. In  (\ref{coulomb}) and  (\ref{quartic}) we have omitted the explicit color dependence, that is discussed in detail in the appendix, where it is also shown how the large $N$ limit considerably simplifies the form of the effective Hamiltonian for the states (\ref{colorless}). The net effect is that, at the leading order in $1/N$, the only surviving interactions are between two neighborhood effective colorless particles in the chains constituting the states (\ref{colorless}). These interactions have exactly the form  (\ref{coulomb}) and (\ref{quartic}).
Thus, the effective Hamiltonian's  matrix elements between two hadronic states with flavors $(i_1,\ldots,i_J)$ and $(j_1,\ldots,j_J)$ are given by (for a comparison with the study of baryons at large $N$, see equations (13) and (15) of ref. \cite{witten})
\begin{eqnarray}\label{ham}
{\cal H}_{i_1\ldots i_J}^{j_1\ldots j_J}& 
=& \sum_{I=1}^J \delta_{i_1}^{j_1}\cdots\delta_{i_{I-1}}^{j_{I-1}}
\delta_{i_{I+2}}^{j_{I+2}}\cdots\delta_{i_{J}}^{j_{J}}\left[  
\delta_{i_I}^{j_I}\delta_{i_{I+1}}^{j_{I+1}}\left(m  - \frac{\nabla_{\vec x_I}^2}{2m} 
- \frac{\lambda}{|\vec{x}_I-\vec{x}_{I+1}|}
\right)\right.
\cr
&&
\left.+ \frac{\lambda}{(2m)^2}\delta^3(\vec x_I-\vec x_{I+1})(\delta_{i_I}^{j_I}\delta_{i_{I+1}}^{j_{I+1}}
-2\delta_{i_I}^{j_{I+1}}\delta_{i_{I+1}}^{j_I})\right]\,.
\end{eqnarray}
These are just the matrix elements, on our states, of the Hamiltonian that includes the mass, kinetic terms and Coulomb interaction (\ref{coulomb}), all diagonal in the flavors, and the point-like interaction (\ref{quartic}), that has a non-trivial flavor structure.

Notice that we do not deal here with any sort of Hartree approximation, where each scalar would 
have experienced an average potential, but instead the nearest neighbor particles 
screen the effect of the rest. Roughly speaking, unlike the mean field theory, 
the energy of a particular scalar is very sensitive to the actual position of its nearest 
relatives. This sensitivity can only be smoothed  in the large $N$ limit.
The above expression, (\ref{ham}), can be rewritten in a more compact form by introducing 
the flavor Hilbert space $({\mathbb C}^2)^{\otimes J}$ with the following permutation 
operator acting on ${\mathbb C}^2\otimes {\mathbb C}^2\,,
P\,: \vert a\rangle \otimes \vert b\rangle \mapsto \vert b\rangle \otimes \vert a\rangle\,,$
\begin{eqnarray}
\label{ha}
{\cal H} = {\cal H}^0+ \frac{2\lambda}{(2m)^2}\sum_{I=1}^J(1-P_{I,I+1})\delta^3(\vec x_I-\vec x_{I+1})\,,
\end{eqnarray}
where we have defined ${\cal H}^0$ as an operator containing just the space dependence 
\begin{eqnarray}\label{spacial}
{\cal H}^0=Jm+\sum_{I=1}^J\left[ -\frac{\nabla_{\vec x_I}^2}{2m} 
- \frac{\lambda}{|\vec{x}_I-\vec{x}_{I+1}|}
- \frac{\lambda}{(2m)^2}
\delta^3(\vec x_I-\vec x_{I+1})\right]\,.\label{hamilto}
\end{eqnarray}
As one might have guessed, in the heavy mass limit, flavor interactions are subdominant. 
But as we depart from this limit they ought to be included. 

Note that in the spin-chain derivation in the ${\cal N}=4$ SYM case \cite{minzar} the whole Hamiltonian is proportional to the $(1-P)$ factor, the latter coming out in its precise form from some crucial cancellations between the gluon exchange terms, the quartic coupling and self-energy and wave-function renormalization terms.
In the present case the $(1-P)$ factor comes out in a different way: there are no self-energy and wave-function renormalization terms and it is only when considering the flavor dependent energy excitations above the spatial excitations given by (\ref{spacial}), that the $(1-P)$ factor appears.

The next step is to find the ground state hadron. 
From the Hamiltonian (\ref{ha}) it is clear that
the flavor 
structure of the ground state is
composed of a single flavor, the same for all the $J$ components. As a consequence the 
action of  $P_{I,I+1}$ on such a state will coincide with the identity operator and allows 
to identify the energy of the ground state as 
given by  ${\cal H}^0$, (\ref{hamilto}). 

Now, our main goal is to find the dependence of the mass spectrum on the flavor degrees of freedom.
As usual in theories in the non-relativistic limit, the spatial $\delta$ factors can be considered as perturbations.
Thus, the flavor dependence is encoded in the Hamiltonian above as a perturbation of the Hamiltonian ${\cal H}^0$, and triggers the splitting in the energy levels of the latter in the flavor Hilbert space.
In other words, we are going to measure the deviation in energy in our hadrons, due to their flavor structure, with respect to a given spatial mass level, the latter being naturally the ``single flavor'' ground state.
Thus,
using standard perturbation theory, 
the many-body wave function for the time independent Schr\"odinger equation 
can
be written as   
\begin{equation}
\vert \Psi\rangle = \vert \phi(\vec x_1,\vec x_2,...,\vec x_J)\rangle \otimes \vert  f\rangle \,,\quad 
\vert f\rangle \in ({\mathbb C}^2)^{\otimes J}\,,
\end{equation}
with the proper normalized particle wave function and 
$\vert \phi(\vec x_1,\vec x_2,...,\vec x_J)\rangle$ referring to the coordinate dependent part of the 
wave function 
for the
ground state obtained from ${\cal H}^0$. Considering the transition between two such states we have
\begin{eqnarray}
\langle\Psi^\prime \vert {\cal H}|\Psi\rangle =  E^0 \delta_{f'f}
+\frac{\lambda}{2m^2}\sum_{I=1}^J \langle
\phi\vert \delta^3(\vec x_I-\vec x_{I+1})\vert \phi\rangle\langle f^\prime \vert (1
-P_{I,I+1})\vert f\rangle \,.
\end{eqnarray}

Due to the symmetry of the ground state, the matrix element $A(\lambda)=
\langle \phi \vert \delta^3(\vec x_I-\vec x_{I+1})\vert \phi \rangle$ will
not depend on the index $I$, thus leading to
\begin{equation}
 \langle\Psi^\prime \vert {\cal H}|\Psi\rangle =E^0 \delta_{f'f}
+ \frac{\lambda A(\lambda)}{2 m^2}\sum_{I=1}^J \langle f^\prime \vert (1-P_{I,I+1})\vert f\rangle \,.
 \label{sandwich}
\end{equation}
The first term on the right hand side of (\ref{sandwich}) is just the renormalized mass of the ground state hadron, that we can write as $E^0=m_{eff}(\lambda)J$ (where $m_{eff}(\lambda)\simeq m+{\cal O}(\lambda)$), and comes from each particle constituting the hadron and their spatial interactions. The second term in (\ref{sandwich}) is the correction to the hadron mass due to its flavor structure. 
We can write the operator giving the flavor dependent corrections to the ground state in the form
\bea\label{spinchain}
\Delta{\cal H}_{\rm red}\equiv {\cal H}_{\rm red}-m_{eff}(\lambda)J=B(\lambda)\sum_{I=1}^J  (1-P_{I,I+1}) \, ,
\eea
where with  the suffix ``red" we mean that the operators act only on the reduced flavor space, once the space part has been fixed to the ground state for the Hamiltonian ${\cal H}^0$ (\ref{hamilto}); we have also introduced an overall factor  $B(\lambda)$ that in the non-relativistic limit we are considering is given by $\lambda A(\lambda)/2 m^2$.
Equation (\ref{spinchain}) explicitly shows that the spectrum for these states is effectively described by a flavor Hamiltonian that
corresponds exactly to the Heisenberg $SU(2)$ spin chain! 

The result (\ref{spinchain}) is relevant for two reasons.
First, the spin-chain is integrable, allowing to calculate the mass spectrum, 
at least in principle, with the Bethe ansatz machinery.
Second and most important, the long wave-length limit of this spin-chain can be described by the same sigma-model that comes from the large $J$ limit of the string sigma-model 
on $S^3$ \cite{k}.
As such, we find agreement at the level of sigma-model in two very different regimes of the theory, the non-relativistic field 
theory one and the string theory one,  up to the undetermined $B(\lambda)$ factor in front of the spin-chain Hamiltonian, to be calculated.
As a preliminary estimate, note that, for dimensional reasons and making use of the hydrogen atom wave function one obtains $B(\lambda)\sim m\lambda^4$. 

In the string description of these hadrons, the classical part of the energy/charge 
relation is given by the above-mentioned sigma-model, with a precise prediction for the overall factor $B(\lambda)$ in the strong coupling limit. 
As we are going to review in the following section, the first $\alpha'$ corrections contribute, for the circular string configuration, just with a overall factor, that we could identify with the mass renormalization.
The latter is of course very different in the field theory and string theory regime, but being an overall factor it does not forbid to compare the remaining structure, namely the sigma-model.
All this suggests the possibility that in the whole $SU(2)$ sector the $\alpha'$ corrections could enter just as overall factors, permitting the matching in general.


\section{The string theory side}

We will now turn our attention to the supergravity side of the duality, discussing the classical and $\alpha'$-corrected energy/charge relation.
In particular we focus again in the four dimensional Witten
model \cite{wym}, studied in \cite{noi4} (we refer to these papers for details). But similar conclusions has been reached within other setups 
\cite{strassler,bcm,sonne} and can be generalized to Witten-like models in any dimension.

The supergravity background of the model contains a four-sphere, whose $SO(5)$ symmetry is the flavor symmetry of five of the six real scalars of the dual gauge theory.
The two natural scales in the theory are given by the field theory string tension $T$ and the lowest Kaluza-Klein mass $m_0^2\,.$
The UV 't Hooft coupling $\lambda$ is related to these quantities by the formula $\lambda=6\pi T / m_0^2$.

The objects dual to the hadrons we have been studying are string states located at the minimal radius \cite{strassler} and spinning in the sphere directions of the metric.
More precisely, they are extended configurations spinning on an $S^3$ inside the $S^4$ with two angular momenta $J_1,\ J_2$ ($J\equiv J_1+J_2$).
In order to deal with the relevant physical region we need to restrict the parameter ranges:
{\sl i)} The ratio of the coupling with $J$ must be small, $\lambda/J \ll 1$.
This assumption
is necessary in order to recover the sigma-model from the spin chain 
\cite{krt}. {\sl ii)} We must demand $J \ll N$ in order to be in the planar regime \cite{minw}.
This requirement is also consistent with that of being far away the giant graviton regime $J \sim N$. 

Concerning the {\it classical} solutions for 
these states, the situation is analogous  to the one of strings spinning in $AdS_5 \times S^5$. For instance, at the minimal radius the metric is just the one for $ R \times S^4\,$ and the results reduce to those reviewed in \cite{Tseytlin:2003ii}.  Furthermore, as already stressed, the {\it classical} sigma-model of a string with angular momenta $J$ moving just on the $S^3$, matches, in the large-$J$ limit, the action of the ``continuous'' limit of the $SU(2)$ spin-chain (\ref{spinchain}) \cite{k}, with an overall  factor given by $B(\lambda)=m_0\lambda^2/9(16\pi)$ \cite{noi4}.
With respect to the FT estimate of the previous section one encounters a mismatch in the power of the coupling, which is $\lambda^4$ in FT and $\lambda^2$ in string theory.
This pattern resembles the one found for the ${\cal N}=4$ theory in the evaluation of the difference, computed in \cite{gubser},  between the conformal dimension and the spin of a particular class of operators;
this difference is proportional to the logarithm of the spin, multiplied by a factor of the coupling which is $\lambda$ in FT and $\sqrt{\lambda}$ in the string calculation. 

Returning to our background, in the $SU(2)$ sector in the large-$J$ limit the circular string configurations satisfy the relation \cite{Tseytlin:2003ii,noi4}
\bea
\label{energy-spincirc}
E_c=m_0J\Big[1+\frac{\lambda^2}{18J^2}\Big(n_1^2\frac{J_{1}}{J}+n_2^2\frac{J_2}{J}\Big)+\ldots\Big]\,,
\eea
where $n_i$ refers to the string winding numbers, which must encode the internal structure of the hadron in the string theory side.  

There are two remarkable facts that distinguish (\ref{energy-spincirc}) from the corresponding $AdS_5 \times S^5$ expression: {\sl i)} The effective coupling constant is $\lambda^2/J^2$ instead of $\lambda/J^2$, a difference that can be traced back to the interplay between the metric with the RR four-form instead of the five-form \cite{strassler}. {\sl ii)} The $\alpha^\prime$ corrections are non-subleading, even in the large-$J$ limit.
At leading order these corrections to the lowest state fit
\begin{equation}
\label{quantum}
E_1 \sim m_0J\Big[-\frac{c}{\lambda}+\ldots\Big]\,,
\end{equation}
being $c$ some positive number \cite{bcm,noi4} ($c=\frac{45}{8}\log{\frac43}$ in this model). Expression (\ref{energy-spincirc})  is functionally equivalent  to
 a particular solution of the sigma-model associated to (\ref{spinchain}) provided the correction in (\ref{quantum}) is interpreted as the stringy equivalent of mass renormalization in FT. 
The comparison of the latter cannot be performed, since 
as we lower the coupling $\lambda$ the semi-classical string
approach breaks down as is signaled by the pole developed in (\ref{quantum}).
Once more, the crucial point is that despite this fact, the remaining structure in (\ref{energy-spincirc}) matches the one in FT.\\

It is straightforward to generalize this discussion to other Witten-like models.
For example, we can obtain a 1+1 YM-theory coupled to KK modes by wrapping D2 branes on a supersymmetry-breaking cycle \cite{wym2d}.
Before wrapping, the D2 theory contains seven massless scalars transforming under $SO(7)$,  and their corresponding fermions.
Without imposing anti-periodic boundary conditions, at low energy we recover the field content of a supersymmetric YM$_{1+1}$ theory.
Instead, with  anti-periodic boundary conditions, and since the fermions acquire a mass proportional to the inverse radius of compactification while the bosons get mass at loop level, we have a 
non-supersymmetric field theory.
It contains, besides the usual YM sector and at the first Kaluza-Klein excitation level, seven massive scalars together with  another massive scalar coming from the field theory living on the circle, and their fermionic cousins, all the fields transforming in the adjoint of $SU(N)$.

{F}rom the gravity side the field components of the ten-dimensional background in the string frame, metric, dilaton field and  a constant six-form field strength, read
\begin{eqnarray}\label{defnsdue}
ds^2&=&\frac{u^{5/2}}{\Lambda^{1/2}} (-dt^2+dx^2 + \Lambda R^2 h(u)d\theta^2)+\frac{\Lambda^{1/2}}{u^{5/2}}\frac{du^2}{h(u)}
+\frac{\Lambda^{1/2}}{u^{1/2}}d\Omega_6^2\ ,\nonumber \\
e^\Phi&=&\frac{\Lambda^{1/4}}{u^{5/4}}\ , \qquad F_6=5\Lambda\omega_6\,,
\end{eqnarray}
with
\eqn{de}{
h(r)=1-\frac{u_0^5}{u^5}\ ,  \qquad R=\frac25 u_0^{-3/2} \ , \qquad \Lambda = 6 \pi^2 g^2_3N\,.
}
The radial coordinate, $u$, ranges inside $[u_0,\infty)$, $\theta$ is an angle, $\Omega_6$ refers to the unit six-sphere and $\omega_6$  to the volume-form of
the transverse $S^6\,.$ The normalization of $F_6$ guarantees that the quantization condition $\int_{S^6}F_6= 32\pi^5N$ is full filled. Moreover we have neglected in the above expressions all the factors of $g_s$ and set $\alpha^\prime=1$\,. Then with this convention the three-dimensional coupling is normalized to the unity, $g_3=1$. 

The running YM coupling can be obtained, as usual, from the DBI action for a D2-brane wrapped on the $\theta$ angle on the background above
\be
{1\over g_{YM}^2(u)} = \frac12\int d\theta
e^{-\Phi}{\sqrt{g_{\theta\theta}}\over g_{00}} =
{2\pi\over5}{\Lambda^{1/2}\over u_0^{3/2}}\sqrt{1-{u_0^5\over u^5}}\ .
\ee
The coupling thus blows up in the IR region $u\rightarrow u_0$ and it
reaches a constant value in the UV $u\rightarrow \infty$, analogously to what happens in the 4d Witten model \cite{wym} . The
corresponding UV value of the 't
Hooft coupling can be easily expressed as
\be
\lambda\equiv g_{YM}^2 N = {5\over 12\pi^3} \Lambda^{1/2} u_0^{3/2}\ ,
\ee
from which it follows that
\be
\lambda m_0 = {5\over 12\pi^3}u_0^3; \qquad {\lambda\over m_0}={5\over
12\pi^3}\Lambda; \qquad \chi\equiv \Bigl({12\pi^3\over5}\Bigr)^{1/3} {\lambda^{1/3}\over
m_0^{2/3}}=\Bigl(\frac{\Lambda}{u_0}\Bigr)^{1/2}\ .
\ee
The two natural scales in the theory are given by the brane tension $T\sim u_0^{5/2}/\Lambda^{1/2}$ and the lowest Kaluza-Klein mass $m_0^2\sim u_0^{3}/\Lambda\,.$
The $\chi$ coupling is given by the natural dimensionless ratio of the UV 't Hooft coupling $\lambda$, that have mass dimension two, and the KK mass scale $m_0$.

Our hadrons are dual to extended strings spinning on a six-sphere with three angular momenta, so formula (\ref{energy-spincirc}) is formally the same but with one charge more.
In order to verify (\ref{quantum}) in this 2d model we can as usual focus on the PP-wave limit of (\ref{defnsdue}), as the first $\alpha'$ correction for the circular string is the same as the zero point energy in that limit  \cite{bcm,noi4}.
Since we are interested in the IR regime, it is convenient to introduce the variable $u^2=(r-r_0)/r_0$ to obtain the expansion to the second order of the metric as
\begin{eqnarray}\label{defnstre}
ds^2&=&\frac{r_0^{5/2}}{\Lambda^{1/2}}(1+\frac52 u^2) (-dt^2+dx^2) + \frac{4\Lambda^{1/2} u^2}{5 r_0^{1/2}}d\theta^2+\frac{4\Lambda^{1/2}}{5r_0^{1/2}}du^2
+\frac{\Lambda^{1/2}}{r_0^{1/2}}(1-\frac{u^2}{2})d\Omega_6^2\ .\nonumber\\
\end{eqnarray}
We can put this metric in a more convenient form by further defining 
$u^2d\theta^2+du^2 \equiv 2 (dY_1^2+ dY_2^2)$ and $(X,\ T) \equiv (m_0x,\ m_0 t)$.
We get
\begin{eqnarray}\label{defns4}
ds^2\sim \chi [(1+ 5 Y_1^2+ 5 Y_2^2) (-dT^2+dX^2) + \frac85 (dY_1^2+dY_2^2)+(1-Y_1^2-Y_2^2)d\Omega_6^2]\ .
\end{eqnarray}
We will use the following parameterization for the six-sphere
\be\label{parsphere}
d\Omega_6^2= d\psi_1^2 + \cos^2{\psi_1} \left[d\psi_2^2 + \cos^2{\psi_2}[d\psi_3^2 + \cos^2{\psi_3}\left( d\psi_4^2 + \cos^2{\psi_4}(d\psi_5^2 + \cos^2{\psi_5}d\psi_6^2)\right) ] \right]\,.
\ee
The null geodesic on the maximal circle is determined by the following  conditions on the coordinates:
$
t=\psi_6\,\,, x^i=u=\psi_i=\theta=0\, (i=1,\ldots,5)\,.
$
As usual we  rescale the coordinates and perform the $L\to\infty$ limit while keeping the lowest Kaluza-Klein mass scale, $m_0$ fixed to obtain the parallel plane-wave metric
\be   
\label{ppw}
ds^2=-4dx^+dx^- - m_0^2\left( v_i^2 +
\frac{15}{4} w{\bar w}\right) dx^+dx^+ + dx dx + dw d{\bar w} + dv_i dv_i\, ,
\ee
where $x^+=t$, $x^-=\frac{L^2}{2}(t-\psi_6/m_0)$, the $v_i$ directions are the rescaled remaining $\psi_i$ coordinates of the six-sphere and the $(w,\bar{w})$ coordinates are the rescaled $r,\theta$ ones.

From (\ref{ppw}) one infers that the spectrum of the string theory on this background is composed by:
{\sl i)} a single massless world-sheet mode, $x$, giving a tower of states of frequencies $\omega_n = n$; {\sl ii)} five massive modes, $v_i$, of frequencies $\omega_n^2 = n^2 + m^2$; {\sl iii)} two $w$ modes of  frequencies $\omega_n^2 = n^2 + \frac{15}{4}m^2$. As for the fermionic modes concerns, taking into account the RR-form in their equations of motion, one gets eight massive modes of frequencies $\omega_n^2 = n^2 + \frac{25}{16}m^2$.
To ascertain the correctness of these results one can check that the Einstein equation for (\ref{defnsdue}) in the pp-wave limit is exactly the same as the mass-matching condition $\sum_{\rm bosons} m_b^2 = \sum_{\rm fermions} m_f^2=\frac{25}{2}m^2$, ensuring finiteness of the theory and absence of Weyl anomaly at one loop.
All in all, these results contribute to the vacuum energy of the theory as
\be
E_1=\frac{m_0}{2m}\sum_{n=-\infty}^{+\infty}\left[n + 5 \sqrt{n^2 + m^2} + 2\sqrt{n^2 + \frac{15}{4}m^2}-8  \sqrt{n^2 + \frac{25}{16}m^2}\right]\ , 
\ee
whose limit for large $m$ gives
\be
E_1 \sim -\frac{m_0 m}{4}[\frac{15}{2}\log{\frac{15}{4}}-25 \log{\frac{5}{4}}]\sim -1.1 m_0 m.
\ee
This is the first $\alpha'$ correction to the classical relation $E_c=m_0J$ and has in fact the expected form (\ref{quantum}), as $m= m_0 p^+= m_0^2 J/L^2=J/\chi$.

An interesting difference with the $4d$ model is in the power of the (dimensionless) coupling that appears in (\ref{energy-spincirc}), that now is $2/3$ instead of $2$. It would be interesting to calculate the corresponding factors of the coupling in the QFT side, but this is an extremely complicated problem, even in this two dimensional system.


\section{Summary}

In this note we have studied the mass spectrum of hadrons composed by many massive particles created by two complex scalar operators in the adjoint representation, in a particular YM theory.
We have shown that in the non-relativistic limit the flavor dependence of the spectrum is encoded in an integrable $SU(2)$ spin-chain.
The simple observation made in this note allows one to attempt to engineer an effective 
theory whose first excited states follow from a spin chain. 
It should be stressed at this point that integrability does not extend to the full 
Hamiltonian obtained from (\ref{denlag}) but only to a sub-sector of operators. One 
can interpret this integrability of the scalar sector as a remnant of the one in ${\cal N}=4$ \cite{minzar}. 
Notice that these states share the same $J^P$ quantum numbers and relate
points in different Chew-Frautschi trajectories. 

The final picture is achieved by comparing the mass obtained in field theory 
with its stringy parallel. 
In both case we have two terms. The first overall term corresponds to the renormalization of the single scalar mass, also due to the spatial interaction with the other scalars in the hadron. 
In the string picture it is included in the value of $m_0$, corrected by the $\alpha'$ contributions (\ref{quantum}), while in QFT it is the $m_{eff}$ factor in (\ref{spinchain}).
The second term is a shift to the mass of the hadron due to its internal flavor structure.
In field theory it is given by the spin-chain in (\ref{spinchain})  and, in the long wave-length limit, by a sigma model.
In string theory, it is given by the same sigma model, showing up as the limit of the classical string sigma model on the three-sphere \cite{k} (expression (\ref{energy-spincirc}) comes from a particular solution).

We expect this matching to be ``universal'' 
and to apply to all the models akin to the one in \cite{wym}, in any dimensions.
Moreover, being the presence of a three-sphere in the IR region of $4d$ confining backgrounds in type IIB, and of the corresponding complex scalars in the dual field theory, common to other models, we expect our results to be true in those settings too.
For example, we could envisage that in all the wrapped-brane models the matching should work as well.  
It would be interesting to check this statement.
Note that the construction does not depend neither on the details of the field 
theory, nor on the compactness or not of its spatial dimensions.
The only ingredients used in the derivation, apart from the non-relativistic limit, are the existence of a confining potential and 
of two complex massive scalars.
This fact reflects the ``universality'' of the dual string theory description of these hadrons, 
that only requires a $R\times S^3$ sigma model in the IR of a confining background.

Finally, let us stress that the power of the coupling in (\ref{energy-spincirc}) is related to the quantity $B(\lambda)$ in (\ref{spinchain}). At this point this constant can be found on numerical grounds or empirically estimated, even though for some simple potentials there is an analytic solution \cite{mcguire}.
It would be very useful to derive it in a rigorous way.

\section*{Acknowledgments}

We are indebted with F. Bigazzi for his participation at different stages of this project. 
We would also like to thank A. Dominguez, E. Lopez, L. Pando-Zayas, J. Russo, W. Troost and especially J. Soto
for very useful discussions.
The research of P.T. was supported by the RyC program. 
This work is partially  supported by the European
Commission RTN Programs MRTN-CT-2004-005104, MRTN-CT-2004-503369,
CYT FPA 2004-04582-C02-01, CIRIT GC 2001SGR-00065.

\section*{Appendix: Color structure and large $N$ limit}

In this appendix we would like to discuss more in detail the color structure entering the interaction terms (\ref{coulomb}) and (\ref{quartic}), and how in the large $N$ limit the effective Hamiltonian takes the simplified chain-like form (\ref{ham}) at the leading order.

To this aim, it is useful to adopt the 't Hooft's double index notation for the elementary (colorful) states forming the multi-particle states (\ref{colorless}). Thus, a single particle state will be indicated synthetically with $|i,\alpha\bar\beta\rangle$ (suppressing the position dependence), where $\alpha\bar\beta$ denote the color indexes (that is related to the base $|a\rangle$ used in section \ref{sec3} by  $|\alpha\bar\beta\rangle=\sum_a (t_a)_{\alpha\bar\beta}|a\rangle$) and we have restricted ourselves to the $SU(2)$ flavor indices $i,j,\ldots$  Thus, if $V_{ii'}^{jj'}=V_{ii',{\rm Coulomb}}^{jj'}+V_{ii',{\rm quartic}}^{jj'}$, the color dependent matrix element of the potential $\hat V$ between two-particle states of the form $|i_1,\alpha_1\bar\beta_1\rangle\otimes  |i_2,\alpha_2\bar\beta_2\rangle$ is given by
\bea
\hat V^{j_1j_2;\gamma_1\bar\delta_1,\gamma_2\bar\delta_2}_{i_1i_2;\alpha_1\bar\beta_1,\alpha_2\bar\beta_2}=\frac{1}{N}\delta^{\gamma_1}_{\alpha_1}\delta^{\bar\delta_1\gamma_2}\delta_{\bar\beta_1\alpha_2}\delta^{\bar\delta_2}_{\bar\beta_2}\times V^{j_1j_2}_{i_1i_2}\ .
\eea

In 't Hooft's double index notation, a colorless state of the kind (\ref{colorless}), again restricting to $SU(2)$ indeces, takes the form
\bea\label{colorless2}
||\psi_1,i_1;\ldots;\psi_J,i_J\rangle\rangle={\cal N}\sum_{\alpha_1,\ldots,\alpha_J}|\psi_1,i_1,\alpha_1\bar\alpha_2;\psi_2,i_2,\alpha_2\bar\alpha_3;\ldots;\psi_J,i_J,\alpha_J\bar\alpha_1\rangle\ ,
\eea
where, in the large $N$ limit, ${\cal N}$ goes like ${\cal N}\sim 1/\sqrt{N^J}$.  
The matrix element of the total effective potential for the states (\ref{colorless2}) is thus schematically given by
\bea
(V_{\rm tot})_{i_1,\ldots,i_J}^{j_1,\ldots,j_J}\!\!\!&=&\!\!\frac{{\cal N}^2}{N}\!\sum_{\alpha_1,\ldots,\alpha_J}\sum_{\beta_1,\ldots,\beta_J}\sum_{1\leq r\leq s\leq J}\delta_{\alpha_1}^{\beta_1}\delta_{\bar\alpha_2}^{\bar\beta_2}\delta_{\alpha_2}^{\beta_2}\cdots\delta_{\alpha_r}^{\beta_r}\delta^{\bar\beta_{r+1}\beta_s}\delta_{\bar\alpha_{r+1}\alpha_s}
\delta_{\alpha_{r+2}}^{\beta_{r+2}}\cdots\delta_{\bar\alpha_{s}}^{\bar\beta_{s}}\delta_{\bar\alpha_{s+1}}^{\bar\beta^{s+1}}\cdots\cr
&&\qquad\qquad\cdots\delta_{\alpha_J}^{\beta_J}\delta_{\bar\alpha_1}^{\bar\beta_1}\delta_{i_1}^{j_1}\cdots\delta_{i_{r-1}}^{j_{r-1}}
\delta_{i_{r+1}}^{j_{r+1}}\cdots\delta_{i_{s-1}}^{j_{s-1}}
\delta_{i_{s+1}}^{j_{s+1}}\cdots\delta_{i_{J}}^{j_{J}}V_{i_ri_s}^{j_rj_s}\ ,
\eea 
and one can easily realize that in the large $N$ limit it becomes
\bea
(V_{\rm tot})_{i_1,\ldots,i_J}^{j_1,\ldots,j_J}&=&\sum_{1\leq r\leq J} \delta_{i_1}^{j_1}\cdots\delta_{i_{r-1}}^{j_{r-1}}
\delta_{i_{r+2}}^{j_{r+2}}\cdots\delta_{i_{J}}^{j_{J}} V_{i_ri_{r+1}}^{j_rj_{r+1}} +{\cal O}(1/N^2)\ . 
\eea
We thus recover a chain-like interaction between the effective colorless particles and thus the resulting effective Hamiltonian in the large $N$-limit is given by (\ref{ham}).


\end{document}